\RequirePackage{fix-cm}
\documentclass[twocolumn]{svjour3}          
\smartqed  
\usepackage{graphicx}
\usepackage{scrtime}
\usepackage[dont-mess-around]{fnpct}

\usepackage{etoolbox}
\usepackage{hyperref}
\newcommand*{\affaddr}[1]{#1} 
\newcommand*{\affmark}[1][*]{\textsuperscript{#1}}
\journalname{Neural Computing and Applications}
\begin{document}

\title{A Dataset and Benchmark for Malaria Life-Cycle Classification in Thin Blood Smear Images
}


\author{%
Qazi Ammar Arshad \protect\affmark[1] \and Mohsen Ali\affmark[1] \and Saeed-ul Hassan\affmark[1] \and Chen Chen\affmark[2] \and Ayisha Imran \affmark[3] \and Ghulam Rasul \affmark[4] \and Waqas Sultani \affmark[1]
}
\authorrunning{Qazi Ammar Arshad \and Mohsen Ali\and Saeed-ul Hassan \and Chen Chen \and Ayisha Imran \and Ghulam Rasul \and Waqas Sultani}

\institute{
            \\
              \email{waqas.sultani@itu.edu.pk}           
          \and
            \at
              \\
\affaddr{\affmark[1]Information Technology University, Lahore, Pakistan}\\
\affaddr{\affmark[2]University of North Carolina at Charlotte, USA}\\
\affaddr{\affmark[3]Chughtai Institute of Pathology, Lahore, Pakistan}\\
\affaddr{\affmark[4]Ittefaq hospital, Lahore, Pakistan}
}
\date{Received: date / Accepted: date}
\maketitle
\begin{abstract} Malaria microscopy, microscopic examination of stained blood slides to detect parasite \textit{Plasmodium}, is considered to be a gold-standard for detecting life-threatening disease malaria. 
Detecting the plasmodium parasite requires a skilled examiner and may take up to  10 to 15 minutes to completely go through the whole slide.
Due to a lack of skilled medical professionals in the underdeveloped or resource deficient regions, many cases go misdiagnosed, which result in unavoidable medical complications.
We propose to complement the medical professionals by creating a deep learning-based method to automatically detect (localize) the plasmodium parasites in the photograph of stained film.
To handle the unbalanced nature of the dataset, we adopt a two-stage approach. Where the first stage is trained to detect blood cells and classify them into just healthy or infected. The second stage is trained to classify each detected cell further into the malaria life-cycle stage. To facilitate the research in machine learning-based malaria microscopy, we introduce a new large scale microscopic image malaria dataset. Thirty-eight thousand cells are tagged from the 345 microscopic images of different Giemsa-stained slides of blood samples.  
Extensive experimentation is performed using different Convolutional Neural Networks on this dataset. Our experiments and analysis reveal that the two-stage approach works better than the one-stage approach for malaria detection. To ensure the usability of our approach, we have also developed a mobile app that will be used by local hospitals for investigation and educational purposes. The dataset, its annotations, and implementation codes will be released upon publication of the paper. 
\end{abstract}
\section{INTRODUCTION}
Out of 400 species of the \textit{Anopheles} mosquito only 60 species carry \textit{plasmodium}, a parasite that causes malaria. Of that, only a female mosquito's bite can transfer the parasite to humans. 
According to the World Health Organization, only in 2018, 228 million cases of malaria occurred worldwide and there were 405,000 deaths globally due to malaria \cite{world2019world}..
Malaria-causing mosquito, Anopheles, is found in tropical and subtropical regions including Africa, Latin, America, and Asia; and adversely affects many resource-deprived regions. 
African continent shares $90\%$ of the deaths caused by malaria, with $68\%$ of the Ethiopian population and $97\%$ of the Democratic Republic of the Congo lives in areas at risk of malaria. \cite{alemu2017performance,mukadi2013external}.

Although rapid testing kits are becoming more common, the standard method of diagnosis of malaria is a microscopic analysis of the stained blood films or slides \cite{abbas2018machine,devi2018hybrid,mukadi2013external}.
Being a low-cost and simple technique, it's a vital tool for detecting both malarial parasites and their density.
Treatment in the early-stage being vital for avoiding medical complications rely heavily on early diagnosis. 
Examination of slides is taxing and difficult since a small portion of the slide is visible through microscope and parasite itself is of small size and available in low-density \cite{world2015methods}”, resulting in 10-15 minutes for a thorough examination.
Adding to the time constraint, low-quality equipment or not sufficiently skilled medical personnel \cite{malera2011research} further adversely affect the quality of malaria microscopy analysis, leading to inaccurate results \cite{ashraf2012developing,wongsrichanalai2007review}. For the successful recovery of a patient, it is vital to diagnose and treat malarial infection early. Named as among diseases of the poor, malaria mostly affects the poorest communities of the world, where preventive measures are not affordable and medical treatment is not readily available.
We propose a method to supplement the medical professionals in increasing the accuracy and decreasing the time required for the examination.
Once the slides are prepared, the images could be taken from either the microscopic camera or using the mobile phone camera through the microscope's eye-piece.
Malaria cells will be detected, counted, and classified by our algorithm. 
The output of our algorithm shown by overlaying results on the image can help medical professionals to analyze decisions made by the algorithm.  With the help of our such technology even a  less trained expert, not surprisingly common in developing countries \cite{ashraf2012developing,wongsrichanalai2007review}, can detect malaria parasites in the field prepared blood slides. 
 
In this paper, we present a two-stage approach for malaria detection and malaria life-cycle-stage classification.
 During the first stage, given the image captured by a microscopic camera or mobile phone, microscopic cells are efficiently segmented using morphological operation and watershed algorithm. This is followed by the extraction of deep convolution features for malaria versus non-malaria classification. Once the malaria cells are detected, in the second stage, we employ another deep classifier for the malaria stage classifications.  
 
 Currently,  no large public microscopic image dataset annotating malaria is available, especially one from the developing country. 
We have collected samples from a local hospital in Lahore, Pakistan, and got them annotated from the expert. 
Our dataset contains P.vivax malaria species in four different life cycle stages including Ring, Schizont, Trophozoite, Gametocyte.
Finally, to make our approach to be used in practical scenarios, we have also developed a mobile app that makes our approach user friendly.
Extensive experiments are conducted to handle the bias introduced due to the class imbalance in the dataset. The experimental results indicate the efficacy of our approach. 
 
 \section{Related Work}
 
The automatic analysis of medical images, i.e., the images containing microscopic cells \cite{AppliedSoftComputingCells,AppliedSoftComputingCells2,abbas2018machine,devi2018hybrid,khashman2012investigation}, MRI \cite{AppliedSoftComputingMR,AppliedSoftComputingMR2}, CT-scans \cite{AppliedSoftComputingCT,AppliedSoftComputingCT2} etc., can have huge impact on accuracy and efficiency of diagnostics. Due to this, several recent works have devised automatic methods for medical image analysis; including the images containing the malaria parasites.

Devi et al. \cite{devi2018hybrid} introduced a hybrid classifier that is a combination of three different classifiers (SVM, kNN, and ANN) and used it for the life-cycle stage classification of malaria cells. Their results show that their proposed hybrid classifier performs better than the individual classifier. Gloria et al. \cite{diaz2009semi} adopted a two-stage approach and used histogram features for the classification of malaria cells. The first stage is used to classify healthy and malaria cells, while the second stage is used to identify the in-between malaria stages (ring, trophozoite, gametocyte, and schizont). Abbas et al.  \cite{abbas2018machine} proposed a color segmentation through an adaptive algorithm of Gaussian mixture model for the classification of healthy and malaria-infected cells in thin blood smear and estimated the level of parasitemia. Khashman .\cite{khashman2012investigation} used convolutional neural networks for the classification of blood cells for microscopic images.  Their neural network model classified the blood cells into three classes i.e., red blood cells, white blood cells, and platelets. Kiskin et al .\cite{kiskin2020bioacoustic} developed an algorithm to detect mosquitoes from acoustic data and showed that audio recording classification through CNN employing wavelet transformations achieve better results as compared to the classifiers that use handcrafted features. 

Di Ruberto et al. \cite{di2002analysis} introduced automatic detection and classification of Giemsa stained blood cells. They employed a morphological operation for cell segmentation and enhanced cell roundness by disk base structuring elements.  Tek et al. \cite{tek2006malaria} used weighted k-nearest neighbor to classify Giemsa stained blood cells. Rao et al. \cite{rao2002automatic} used mathematical morphological operation on Giemsa stained blood slides to analyze the different life cycle stages of malaria species. Sio et al. \cite{sio2007malariacount} introduced a software named MalariaCount that counts the number of infected blood cells. They used binary morphology to classify between malaria vs healthy cells.  Their experimental results on P. falciparum-infected blood slides demonstrated the automatic calculation of the value of parasitemia. However, their approach does not address the classification of different plasmodium species and the life cycle stage of specific malaria species. Tek et al. \cite{tek2010parasite} presented a framework that used a color normalization technique to improve the segmentation of blood cells. After that they extracted cells from these slides, classify them as infected or healthy and also identify the species and life cycle stage of plasmodium. Kumarasamy et al. \cite{kumarasamy2011robust}  proposed a unique solution for malaria species and stage classification.  A radial basis Support Vector Machine (SVM) was trained on binary morphology and texture features to identify the infection stage of the parasite. Linder et al. \cite{linder2014malaria} used SVM for malaria-infected cell identification in Giemsa stained thin blood smear. Parasite candidate regions were selected based on their color and size. After features extraction (SIFT, local binary pattern, and local contrast), an SVM is trained on these features for parasite detection. Bhowmick et al. \cite{bhowmick2012computer} proposed to use the watershed algorithm for cell segmentation on scanning electron microscopic images captured at 2000X magnification. After that, they extracted different geometric features to train a multi-layer neural network for the classification task. Fatima et al. \cite{fatima2020automatic} used adaptive threshold and mathematical morphological algorithms for malaria parasite detection in thin blood films microscopic images. Similarly, Molina et al. \cite{molina2020sequential} designed a sequential classification model in which they have combined three different classification modules. The first module used SVM and the second and third modules used Linear Discriminant Analysis (LDA) for classification tasks.  

Due to the resurgence of deep convolutional neural networks, recent years have witnessed significant progress in different areas of computer vision, including malaria detection
Bibin et al. \cite{bibin2017malaria} introduced a deep learning-based binary classifier for the classification of infected vs non-infected cells using color and texture features and a deep belief network (DBN). A hybrid model for malaria-infected blood cell classification was introduced by Devi et al. \cite{devi2018malaria}. They employed a combination of three classifiers (SVM, k-NN, and Naive Bayes) and also introduced an optimal feature-set which is a combination of some existing and new features.   
Authors in \cite{vijayalakshmi2019deep} introduced a novel deep model in which they replaced the last layer of the VGG network with SVM (named as VGG-SVM) and used it for the classification of plasmodium falciparum-infected blood cells.

Similar to malaria detection in a thin blood smear, several approaches have been proposed for automatic malaria detection in a thick blood smear. Mehanian et al. \cite{mehanian2017computer} used Convolutional Neural Network (CNN) for the detection of the malaria parasite in the thick blood smear with sufficient accuracy.   
Salamah et al. \cite{salamah2019robust} purpose thick blood smear segmentation technique using intensity slicing and morphological operation for malaria parasite detection of microscopic images. The proposed method is robust to noise, artifacts, and intensive variation in blood slides. Yang et al. \cite{yang2019deep} introduced a new publicly available thick blood smear dataset containing 1819 from 150 patients. Classification of malaria parasites is performed in two stages. The first stage uses intensity-based Iterative Global Minimum Screening (IGMS) to generate the parasite candidates and in the second stage, customized CNN is used to classify the candidate region as parasite or background. 

In contrast to the above approach, we presented a new indigenously collected dataset containing microscopic images of malaria slides using a thin smear and put forward a two-stage approach for accurate cell classification. Also, we provide a mobile app and an educational use-case to make our approach user-friendly.
\begin{figure*}
    \centering
    \includegraphics[width=1\textwidth]{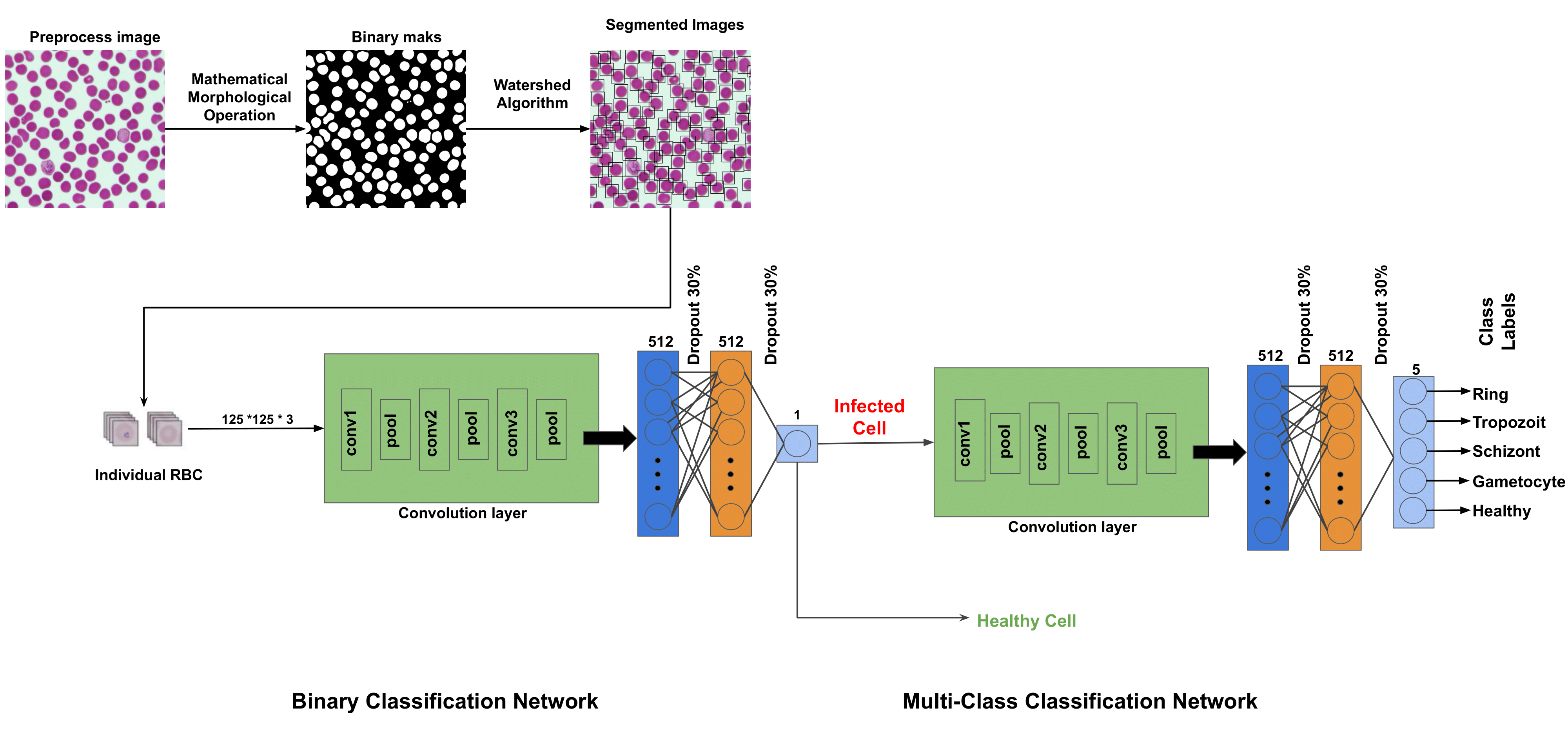}
    \caption{The flow diagram of the proposed malaria detection approach. Processes start with an image captured with a mobile phone/microscopic camera followed by the pre-processing binary mask generation. After segmenting the cells using the watershed algorithm, individual cells are fed into our binary classification network which separates healthy and infected cells. Finally, the infected cells are fed into another multi-class classification model for life cycle-stage detection.}
    \label{fig:solution}
\end{figure*} 
  
\section{Proposed Malaria Detection Method}
Our goal is to detect and localize malaria-infected cells in images that are captured using either a mobile phone camera or through a camera attached with a digital microscope. 
Our proposed approach to tackle this challenging problem is based on five observations: 
(1) since the purpose is to facilitate doctor on time, the proposed method should be efficient and must have high classification accuracy,
(2) in practical scenarios, sometimes doctors just want to know the presence or absence of malaria, while in some other critical situations, they are interested to find out actual malaria stage, therefore two-stage approach which serves both purposes would be preferable,
(3) to make the medical professional comfortable with the method, the approach should be trained and tested on the locally acquire microscopic data,
(4) a user-friendly interface to employ and interpret the output of the algorithm,
(5) finally, to make new medical practitioners more comfortable with computer vision-based diagnostics, the approach should help them in their educational understanding of the problem. 
To make our paper self-contained and be reproducible, we describe details of our approach in the following sections. Specifically, we discuss cell segmentation and the details of the two-stage approach for malaria-infected cell detection. Finally, we discuss the experimental results, our mobile app, and the application of our approach for education purposes.

\subsection{Cropping the area of interest}
\begin{figure}[b]
    \includegraphics[width=0.49\textwidth]{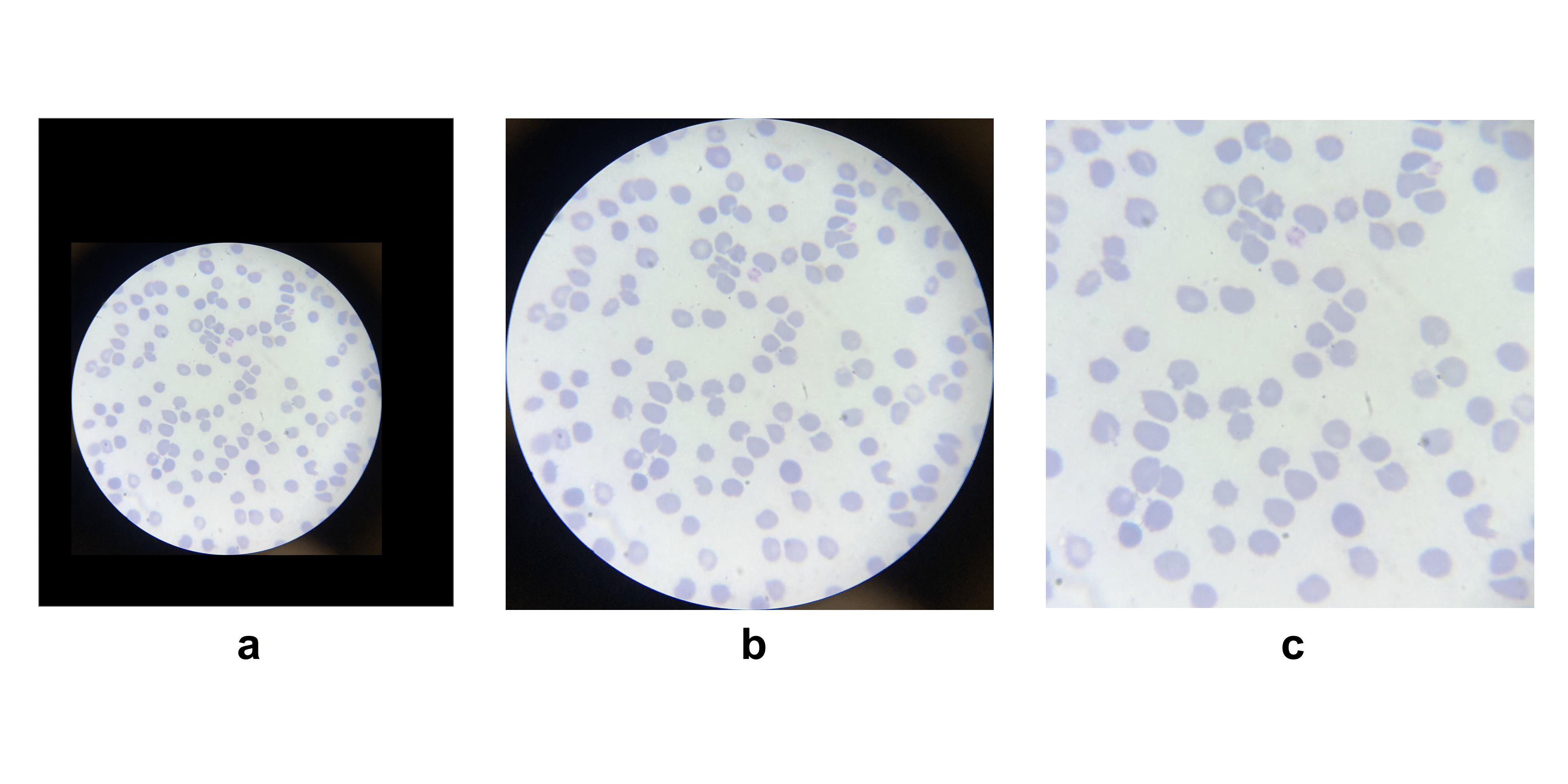}
    \caption{Process of removing unwanted darker regions images captured from microscope through mobile phone: a) shows the unwanted darker region around the original image captured with a mobile phone mounted over an optical microscope, b) shows the image after the first pre-processing step in which image is cropped by finding the largest on the original image, c) represents the image after applying automatic thresholding method of preprocessing to completely remove the darker region from the image.
} \label{fig:black_region_removal}
\end{figure}

Using the mobile phone camera to perform microscopic analysis introduces various artifacts.
One of them is the dark region surrounding the region of interest, observed when viewing the slide through the mobile camera Fig. \ref{fig:black_region_removal}.a. Due to the static focal length in most mobile phones, this could not be removed by camera settings. These darker shaded regions represent the background and could affect the performance of the segmentation method.
Using the basic image processing techniques, these darker regions are removed. 
The first image is converted to the grayscale image and contrast is enhanced by applying the histogram equalization. 
The resultant image is converted into the binary image by a threshold. The threshold is picked using the Otsu method. The original image and crop darker region according to the largest contour as shown in Fig \ref{fig:black_region_removal}.a and Fig \ref{fig:black_region_removal}.b respectively.
 Secondly, we count the ratio of darker to non-darker pixels in both row and column and crop the image from the locations where this ratio goes below a certain threshold. Fig \ref{fig:black_region_removal}.c shows the result of the final image. 

\begin{figure}[b]
\includegraphics[width=0.49\textwidth]{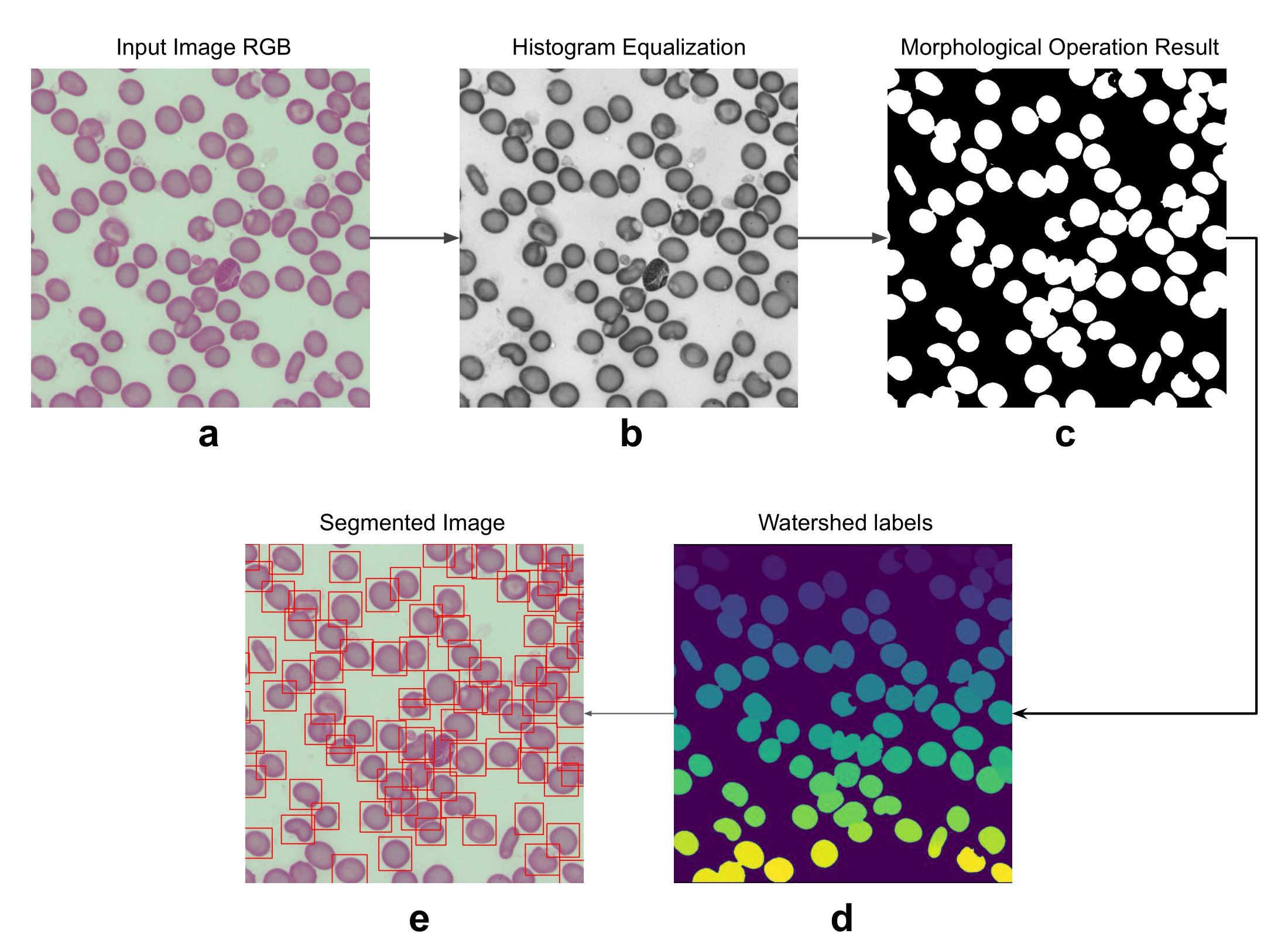}
    \caption{a) shows the original image captured with a mobile phone/microscopic camera, (b) represents histogram equalization is applied to the gray image to enhance contrast, (c) shows a binary mask of the image after applying a set of morphological operations and, (d) demonstrate labels generated using the watershed algorithm. Finally, (e) shows segmentation results indicating red blood cells in a thin blood smear image.
}
    \label{fig:loclization}
\end{figure}

\subsection{Cell Localization}\label{sec:Cell Localization}
An important step for the malaria cell analysis is to identify and localize the cells in the microscopic image, which is predicting the bounding box around all of the cells in the image.
Microscopic cell localization is performed employing the combination of mathematical morphological operations and watershed algorithm. 
Specifically we perform following steps to localize (Fig. \ref{fig:loclization}).
 
\begin{figure}[!ht]
    \includegraphics[width=0.5\textwidth]{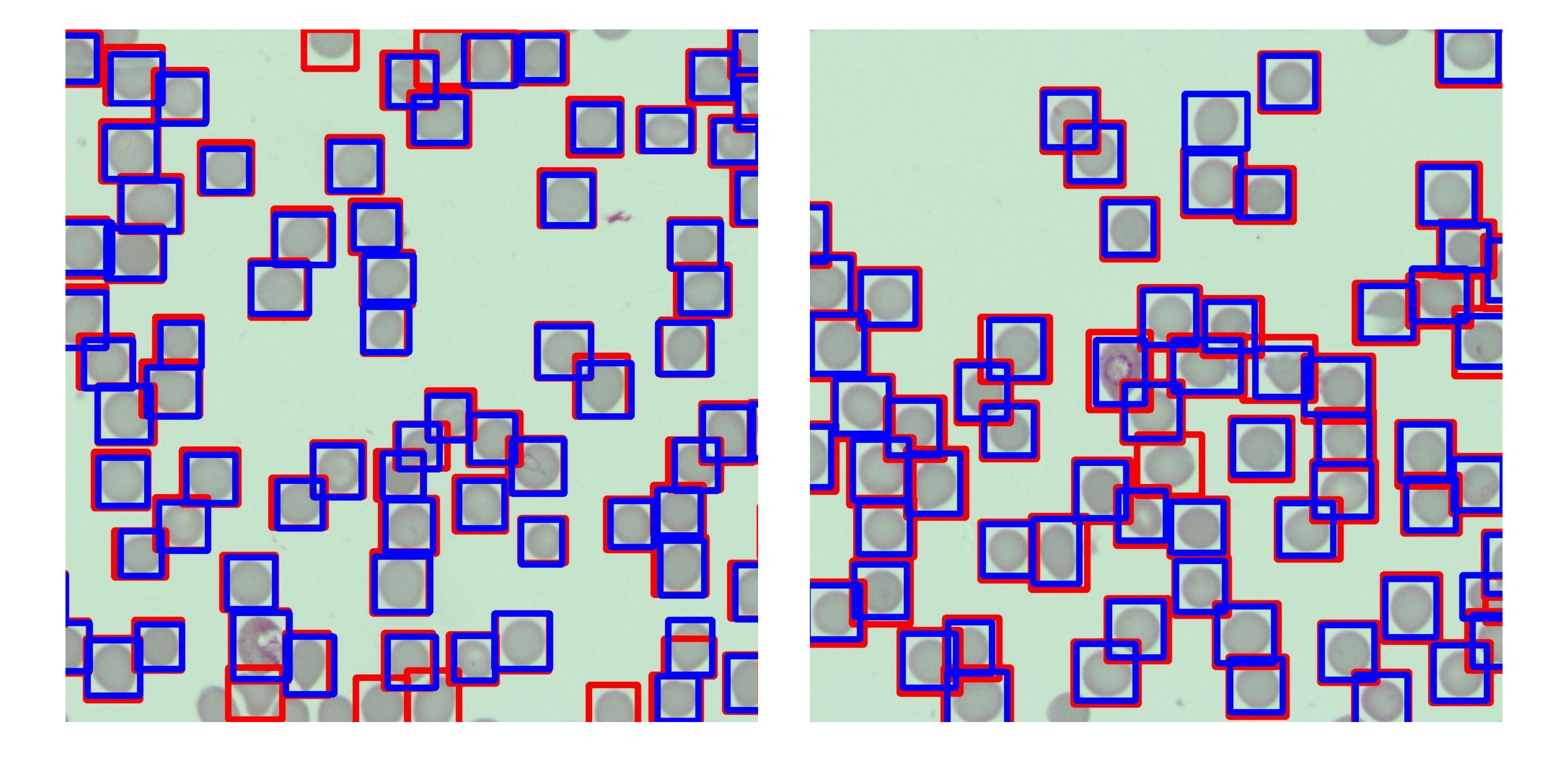}
    \caption{Result of cell localization on IML-Malaria dataset. The blue and red boxes show the ground truth predicted cell bounding boxes respectively.}
    \label{fig:localization_results}
\end{figure}

\begin{enumerate}
  \item \textbf{Binarization:} We first convert the input image into grayscale and apply histogram equalization on the image to enhance the contrast of blood cells from the background. After that, we divide our input image into sub-parts and then apply otsu thresholding \cite{otsu1979threshold} to get a binary mask of these sub-images. In some cases due to non-uniform staining of peripheral blood slides, light illumination may not be the same in the whole image, therefore, this division helps in getting different thresholds for foreground and background in the different regions of the same image. After this, these sub-images are combined to get a binary mask.

  \item  \textbf{Noise Removal:} To enhance the cell structure, we applied two mathematical morphological operations. First, we apply an opening operation to remove small artifacts from the binary image. Second, erosion operation is performed to separate the clustered cells from each other. Figure. \ref{fig:loclization}.c shows the binary mask generated after otsu thresholding and morphological operations.
  
  \item  \textbf{Segmentation:} The watershed algorithm \cite{roerdink2000watershed} is applied on binary masks to generate unique labels for each cell. Finally, we estimate contour on these labels to get the individual cell localization (Figure. \ref{fig:loclization}.d and Figure. \ref{fig:loclization}.e).  A few cell localization results are shown in Figure \ref{fig:localization_results}.
\end{enumerate}

\subsection{Cell's infection-stage Recognition}
Each localized cell is sent to the classification module, which labels each detected cell according to the stage of infection. Two strategies are adopted for the classification module, in the first one, we train a single-stage classification to predict all the labels including whether the cell is healthy or not. 
This strategy, although reasonable, is adversely affected by the imbalanced distribution of healthy and non-healthy samples of the cells. In the second one,
a two-stage setup is introduced to handle this issue. Below we detail both designs. 
\newline\newline
\textbf{Single-stage Classification  (SSC):} A multi-label classification network is trained to classify the cells extracted from the localization step. 
The network consist of a Convolutional Neural Network-based feature extractor is employed to extract the features, which are input to fully-connected layers. The output layers infer the probability of the cell belonging to four malarial life-stage classes i.e.,  ring, trophozoite, schizont, gametocyte, and the healthy class. 
To discover the convolution neural network architecture that performs best for our problem,  we have used  VGG16, VGG19 \cite{simonyan2014very}, ResNet50v2 \cite{he2016deep}, DenseNet169, DenseNet201 \cite{huang2017densely} and the architecture proposed in Rajaraman et al.\cite{rajaraman2018pre}.
All these networks are first trained on imageNet dataset \cite{deng2009imagenet} and then fine-tuned on our dataset.   

Normally microscopic slides dataset contains a large number of healthy cells and very few infected cells. 
Due to high-class imbalance, the learning process is biased towards predicting the health cells correctly rather than differentiating between the stages. 
To counter the effects of class imbalance, we design a cascade structure consisting of two stages, as described below. \newline\newline
\noindent\textbf{Two-stage Classification  (TSC):}
To address the limitations of SSC, we propose a two-stage class classification approach. 
The first stage performs binary classification which classifies cells as healthy and malarial infected.
After that, the cells that are indicated as malaria are further fed into a multi-class classification network to identify the life cycle stage of infected cells.
To handle the situation, where health cells miss-classified as infected, the second stage also predicts healthy as one of the possible labels. 
Figure. \ref{fig:solution} shows the architecture of our two-stage method that is a combination of two different networks.

Similar to SSC,  different CNN models, VGG16, VGG19, ResNet50v2, DenseNet169 and DenseNet201 and the architecture proposed by Rajaraman et al.\cite{rajaraman2018pre} are used in binary classification (first stage).
For the training of the second stage, a subset of training data is generated that contains all malaria classes and randomly sampled healthy cells, such that the training data is balanced.
From the many different versions of the first stage, we select the best binary classifier to classify the cells into malaria and then use the second stage to further classify the malaria cells into their different types. 
Our experiments demonstrate that the two-stage approach outperforms the one-stage approach.

\begin{table*}

\label{table:dataset_comparison}
\begin{tabular}{c|c|c|c|c|c|c}
\hline\noalign{\smallskip}
Malaria Dataset & Malaria Species & \multicolumn{1}{c|}{\begin{tabular}[c]{@{}c@{}}Stage \\ Classification\end{tabular}} & \multicolumn{1}{c|}{\begin{tabular}[c]{@{}c@{}}Classification \\ Labels\end{tabular}} &  \multicolumn{1}{c|}{\begin{tabular}[c]{@{}c@{}}Number of \\ annotated cells\end{tabular}} & 
\multicolumn{1}{c|}{\begin{tabular}[c]{@{}c@{}}BBX \\ Localization\end{tabular}} &  
 \multicolumn{1}{c|}{\begin{tabular}[c]{@{}c@{}}Slide \\ Images\end{tabular}}
   \\ \hline
\noalign{\smallskip}\hline\noalign{\smallskip}
Segmented-Malaria \cite{rajaraman2018pre} & P.f & No & Binary & $27,558$ & No & No  \\
MaMic Image Database \cite{linder2014malaria}   & P.f & No & Binary & $16,991$ & No & Yes \\
\hline
P.vivax (malaria)\cite{BBBC041v1} & P.v & Yes & MultiClass & $80,000$ & Yes & Yes\\
malaria-655\cite{boray2010} & P.f, P.v ,P.m, P.o & Yes & Multiclass & 4363 & No & Yes \\
MP-IDB\cite{loddo2018mp} & P.f, P.v ,P.m, P.o & Yes & Multiclass & 840 & No & Yes \\
IML-Malaria Dataset   & P.v & Yes & MultiClass & $38,449$  & Yes & Yes \\

\noalign{\smallskip}\hline

\end{tabular}
\caption{A comparison of different malaria datasets. Binary classification labeled datasets only describe cells as healthy or malaria while multi-class datasets provide labels of different life cycle stages of plasmodium. whereas P.f represents Plasmodium falciparum, P.v represents Plasmodium vivax, P.m represents Plasmodium malariae and P.o represents Plasmodium ovale.}
\end{table*}


    






\section{Datasets for Automatic Detection of Malaria-parasite}
Machine learning models in general and deep learning in specific are highly dependent upon good quality annotated datasets of reasonable size. 
We found currently available datasets lacking and hence collected a large one to help facilitate the research in this important direction. 
Before describing our malaria dataset, we brieﬂy review the existing microscopic blood smear dataset for malaria detection.
\newline\newline\textbf{Segmented-Malaria} \cite{rajaraman2018pre}: This dataset contains $27,558$ individual cell images of P. falciparum which are manually annotated by trained experts. This dataset contains an equal number of healthy and infected cells where the images of thin blood films stained with Giemsa are captured with a mobile phone mounted on a microscope. The dataset contains individual images of blood cells which limits its usage in the real-world application where more focus is on malarial cell detection and localization.  
\newline\newline\textbf{MaMic Image Database} \cite{linder2014malaria}: This dataset contains images of blood slides that are used to identify P. falciparum. Images are captured with a microscopic camera that scans the whole slide and captured around 549 images of a single slide. Every single image has 1280 $\times$ 1024 pixels resolution. After that,  these images are stitched together and compressed into ECW (Enhanced Compressed Wavelet) format. These ECW file formats are uploaded to an image management platform server WebMicroscope \footnote{http://demo.webmicroscope.net/SlideCollections} that provide a whole-slide view.  For analysis, these images can be download as 1500 $\times$ 1500 pixels tiles in PNG format. The annotation task of blood cells is done by a trained expert that labels the cells as certain parasites or uncertain parasites.
This dataset is relatively small consisting of only $16,991$ blood cells in total. 
\newline\newline\textbf{P.vivax (malaria)} \cite{BBBC041v1}: This dataset consists of 1364 images of thin blood smear stained with Gamesa. The dataset contains around $80,000$ blood cells. The annotation contains cell-level bounding boxes annotations of the healthy and infected cell. Also, the dataset contains annotations of the cells at different life cycle stages including ring, trophozoite, schizont, and gametocyte
\newline\newline\textbf{Malaria-655} \cite{boray2010}: This dataset contains 655 images of nine different thin blood slides that are prepared separately from each other at different time. To prepare their dataset, the blood smear is fixed on the slide with methanol and stained with Giemsa stain. This dataset includes images of four different species of malaria including P.falciparum, P.vivax, P.ovale, and P.malariae. SP200 research microscope with a mounted Canon A60 camera is used for capturing images. 
Images are captured at a 100x objective lens. The images in this dataset are of 1600 $\times$ 1200 pixels resolution. Data is annotated by the trained experts and the annotation contains 669 parasite images and 3431 healthy cell images.
This dataset, however, has only a dot at the center of the cell to represent its location instead of the bounding box. 
This information is not enough to train the object detection algorithms to extract the cells for further processing. 
\newline\newline
\textbf{MP-IDB} \cite{loddo2018mp}: This dataset contains four different genus of malaria parasites including P.vivax, P.falciparum, P.ovale and P. malariae. 
Each genus in the dataset has four distinct life cycle stages of plasmodium. 
Dataset is acquired at 100x magnification with Leica DM2000 optical laboratory microscope coupled with a built-in camera. 
It consists of only 229 images with a 2592$\times$1944 pixels resolution that represents four genus of plasmodium. 
Each image has at least one stage of the malaria parasite in it. It has $48,000$ blood cells and images are labeled by expert pathologists. 
They provide image-level labels that indicate the presence of a malaria parasite and the name of the image contains the information of the life cycle stage of the parasite in a specific image. 
\newline\newline\textbf{Limitations of existing dataset:}
Each of the above-mentioned datasets has its limitations: Segmented-Malaria \cite{rajaraman2018pre} dataset contains individual images of blood cell without image level labels or information. MaMic image database \cite{linder2014malaria} dataset contains a small number of blood cells. P. vivax (malaria) \cite{BBBC041v1} is collected from a developed world and does not necessarily represents our indigenous hematology, Malaria-655 \cite{boray2010} lacks bounding box level annotations and similarly MP-IDB \cite{loddo2018mp} only provides  image-level labels. Table \ref{table:dataset_comparison} demonstrates the detailed comparisons between existing and the proposed dataset.\newline\newline\textbf{Our Dataset--IML-Malaria:}

\begin{figure}[t]
    \includegraphics[width=0.5\textwidth]{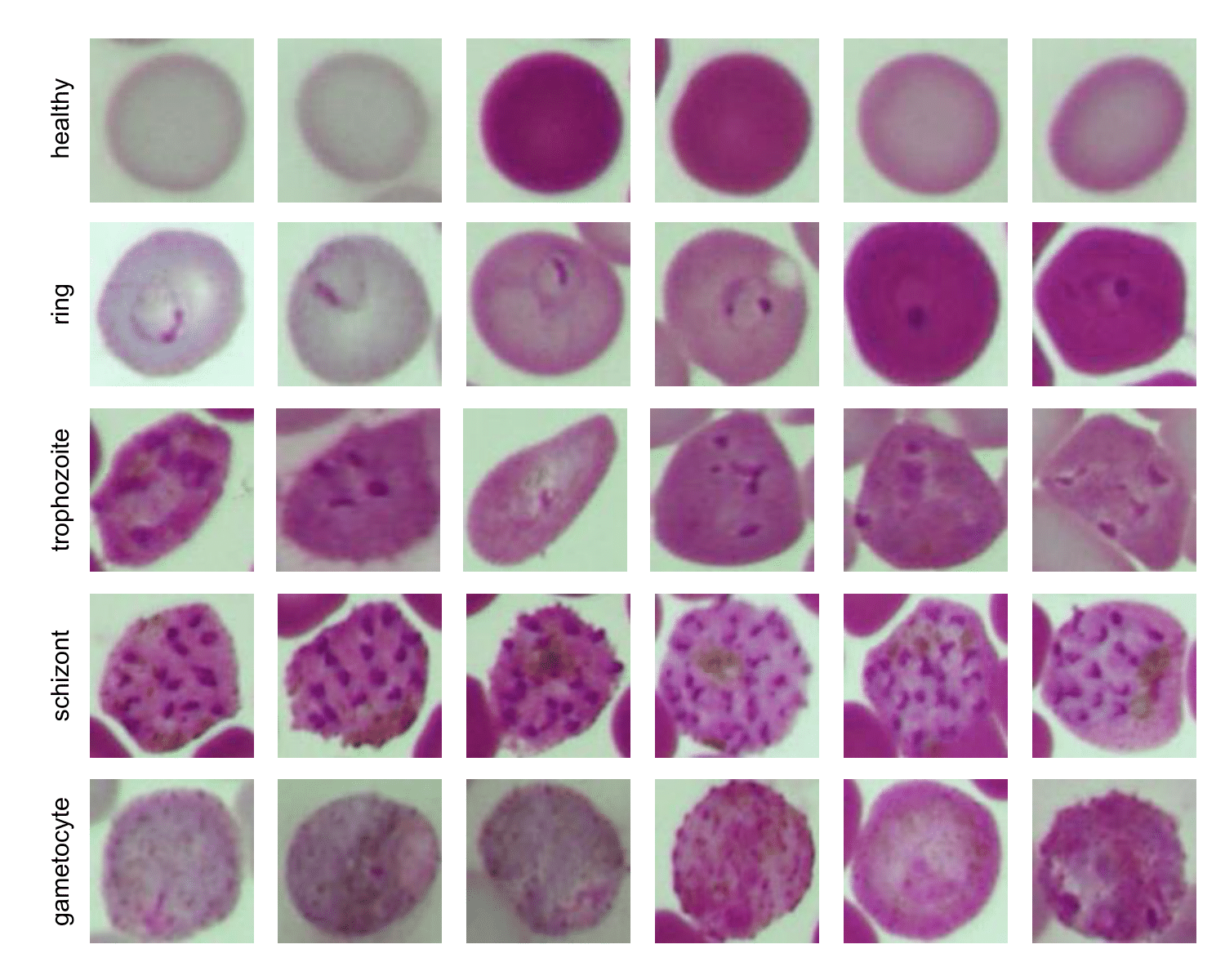}
    \caption{This figure shows different stage of malaria including ring, trophozoite, schizont and gametocyte in IML-Malaria dataset. The top row shows healthy cells. Please see Section 1.1 for the details of different malaria stages.}
    \label{fig:all_malaria_stages}
\end{figure}

\begin{figure*}
\centering
\includegraphics[width=1\textwidth]{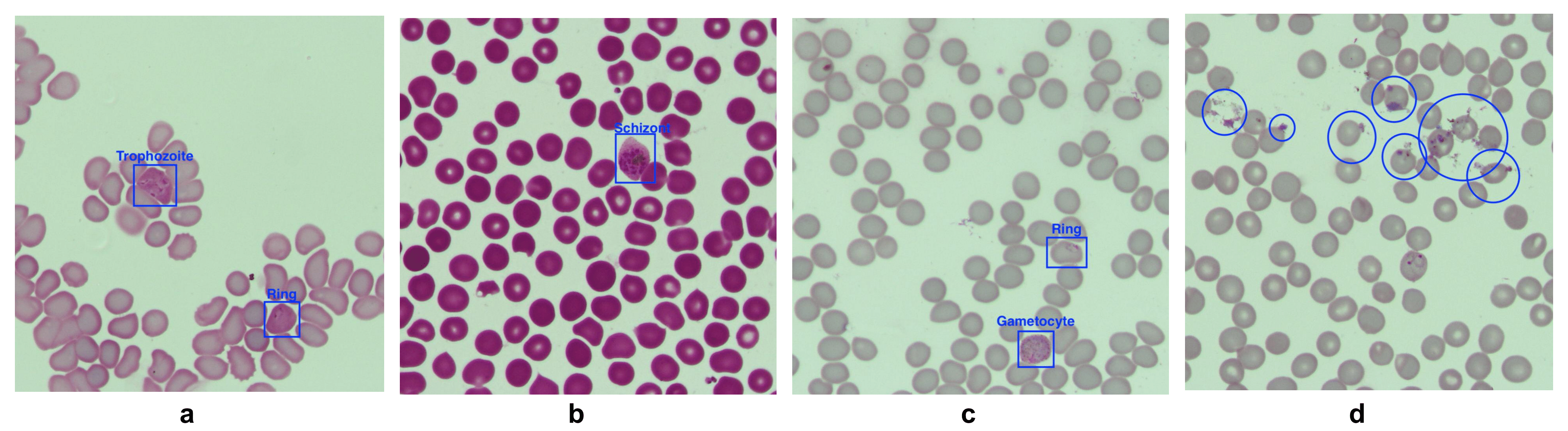}
\caption{Examples of images in IML-Malaria dataset. Slide images look different to each others, due to the effect of different staining conditions. The malaria stages shown as a) Trophozoite and Ring stage b) Schizont stage c) Ring and Gametocyte stage. Blue circles in (d) show different regions containing artifacts in the blood slide.}
    \label{fig:annotations}
\end{figure*}

We collected blood samples from malaria-infected people living in the province of Punjab, Pakistan.
Pakistan being an agricultural country, having warm and humid weather, and a large irrigation network is among regions with a high risk of malarial infection \cite{kakar2010malaria}.
 Like many developing countries, malaria control efforts have been hampered due to a lack of expensive laboratory instruments, and, more importantly, skilled technicians and doctors specifically in the remote areas \cite{qureshi2019occurrence}.
Every year, on average, 500,000 confirmed cases of malarial infection and 50,000 deaths are caused by it are reported in Pakistan \cite{khan2019malaria}. The reasons are agriculture practices, monsoon season rains and vast irrigation network \cite{kakar2010malaria}. 
Another reason for malaria infection in Pakistan is frequent floods \cite{qureshi2019occurrence}.  Two major species causing malaria in Pakistan are P.vivax and P.falciparum. The extensive research done by Qureshi et al. \cite{qureshi2019occurrence} shows that 81.3\% cases were caused by P. vivax, 14.7\% by P. falciparum, and the remaining 4\% cases are caused by a mixed infection in Pakistan. 
Pakistan is facing many challenges in malaria control due to misdiagnosis, lack of expensive laboratory instruments, and, more importantly, skilled people specifically in remote areas \cite{qureshi2019occurrence}.


We have collected a new malaria dataset that is captured with an XSZ-107 series microscope. 
Globally 49\%\ cases of malaria \cite{qureshi2019occurrence,alias2014spatial} are reported due to P.vivax, therefore, we concentrate on collecting images of microscopic slides infected by genus of plasmodium.
Blood smears were prepared by spreading thin blood film on the slide and were left to dry.
After fixing the blood smear through methanol, the slide is stained using Giemsa solution for 15 - 20 minutes. 
Finally, the cover-slips are permanently fixed on the slide to keep them safe. Our thin-film blood slides are prepared by trained experts in labs. Images are captured with the help of a camera mounted on a microscope at 100x \footnote{The actual magnification is 100 $\times$ 10 (eye-piece)} objective magnification. Images are annotated and captured by an expert hematologist, at a local institute. 
The dataset contains 345 images consisting of $111$ blood cells on average, both healthy and ones infected by the parasite.
We annotate the infected cells into classes per life-cycle stage, that is, Ring, Trophozoite, Schizont, and  Gametocyte as shown in Figure. \ref{fig:all_malaria_stages}.
In addition to this bounding box for each cell is also annotated. 
It must be noted that trophozoite and schizont are very rare \cite{loddo2018mp} and so they are present in very low numbers in our data collection as well.  Comparison between different datasets are shown in Table \ref{table:dataset_comparison}. Figure  \ref{fig:annotations} some examples of images in our dataset.

\section{Experiments and Results}

We evaluate our proposed approach on two levels of microscopic analysis, the first one is cell localization and another one is stage classification.

\subsection{Cell Localization}
As our CNN are trained on individual cells so we first need to localize blood cells for classification. As mentioned in section \ref{sec:Cell Localization}, we apply the watershed segmentation algorithm with a combination of morphological operations to localize cells in the blood slide image. The cell is considered localized (true positive) if the intersection over union (IOU) between predicted and ground truth cell bounding box is greater than 50\% otherwise the predicted box is considered false positive. Similarly, the cells which are not detected by the algorithm are considered as a false negative. Table.\ref{table:localization} shows the results of cell localization on our dataset. From the table, it can be seen that 34,367 blood cells are accurately detected with 91.14\% $F_1$-Score. 

\begin{table}[h!]\centering
\begin{tabular}{|c|c|c|c|c|c|}
\hline
Dataset                                                & TP    & FP   & P   & R    & $F_1$-score \\ \hline
\begin{tabular}[c]{@{}c@{}}IML-\\ Malaria\end{tabular} & 34367 & 4082 & 0.94 & 0.89 & 0.91     \\ \hline
\end{tabular}
\caption{Localization results on IML-Malaria dataset. TP, FP, P and R represents true positive, false positive, precision and recall respectively.}
\label{table:localization} 
\end{table}

\subsection{Malarial life-cycle stage Classification}

\begin{figure}[h!]
\includegraphics[width=0.48\textwidth]{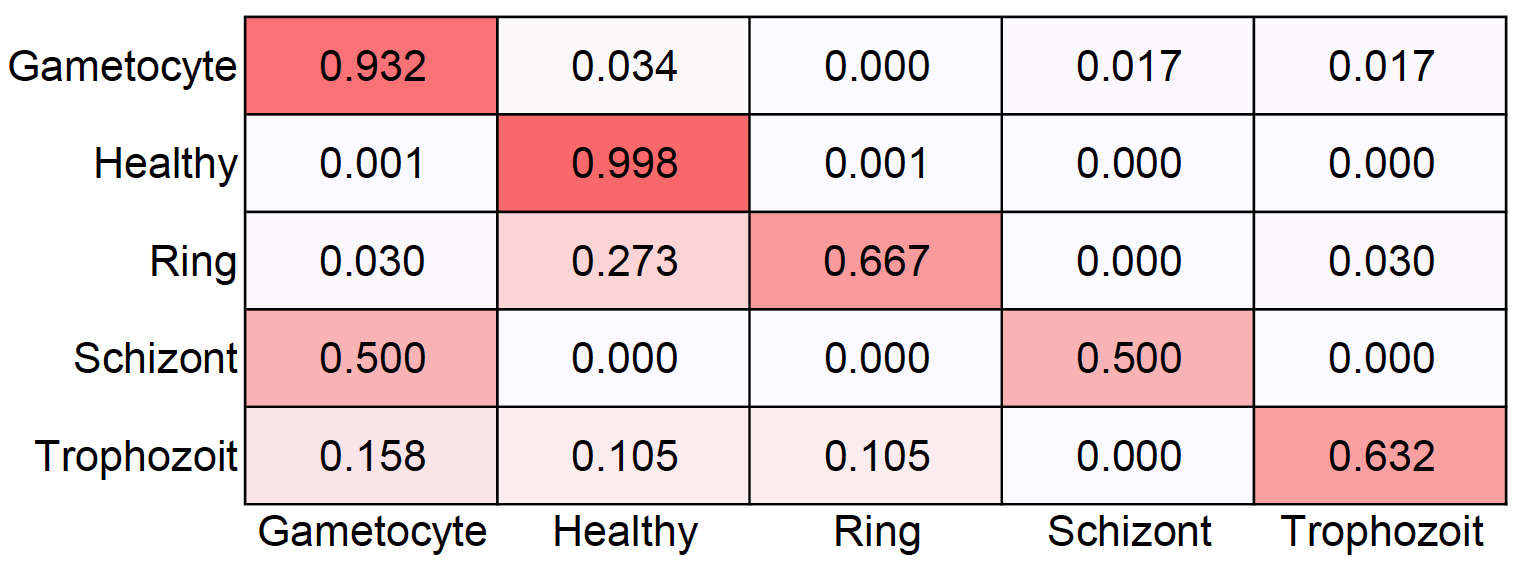}
    \caption{Single stage multi-class classification confusion matrix of DenseNet201 that outperforms in single stage setting on test set of IML-Malaria dataset.}
    \label{fig:single_stage_confusion_matrix}
\end{figure}

 Plasmodium parasites have different life-cycle phases including ring, trophozoite, schizont, and gametocyte stage. The life cycle stage helps in measuring the degree of the parasite inside the patient's blood. 
Our  IML-Malaria dataset is similarly labeled, including the label healthy for the cells without infection. 
We randomly divide the dataset into training (70\%), testing (20\%), and validation (10\%) set.  
All the base-networks used, were pre-trained on imageNet \cite{deng2009imagenet} dataset before plugged into our pipeline and fine-tuned on the malaria microscopy task. 
In our experiment we use the same setting for all the networks (Figure. \ref{fig:solution}), i.e., using the convolution layer of these state of the art networks and then adding two fully connected layers at the end of each network.  
Accuracy of trained single SSC model and TSC model over the testing dataset are given in
Table.\ref{table:single_stage_classifcation} and Table.\ref{table:2stage_classification}. 
To compute the average accuracy,  we first compute class-wise accuracy and then divide the sum of these accuracies by the number of classes. 
Below we discuss the performance of SSC and TSC separately. 
\newline\newline
\noindent\textbf{Single Stage Classification (SSC) Results :}
Results of SSC with different backbone networks are compared in 
Table.\ref{table:single_stage_classifcation}.
CNN proposed by \cite{rajaraman2018pre} shows the average accuracy of 41\%, whereas DenseNet201 outperforms with an average accuracy of 74\%. 
Although the average accuracy appears reasonable, the confusion matrix, Figure. \ref{fig:single_stage_confusion_matrix}, indicates that the results are skewed by the healthy cells.

\begin{table}[h!]\centering
\begin{tabular}{|l|l|l|}
\hline
                     & Avg. Accuracy   & $F_1$-score            \\ \hline
Rajaraman et al. \cite{rajaraman2018pre}            & 46.1\%          & 51.26\%           \\ \hline
VGG-16              & 55.35\%          & 52.40\%           \\ \hline
VGG-19              & 70.44\%          & 76.48\%           \\ \hline
ResNet50v2           & 72.86\%          & 76.44\%           \\ \hline
DenseNet169          & 73.91\%          & 77.62\%            \\ \hline
\textbf{DenseNet201} & \textbf{74.56\%} & \textbf{75.52\%}  \\ \hline
\end{tabular}
\caption{Single Stage multi-class classification results
  on IML-Malaria dataset. Among different CNN models used, DenseNet201 has the highest classification accuracy.}  
  \label{table:single_stage_classifcation}
\end{table}

\begin{figure}[h!]
\includegraphics[width=0.48\textwidth]{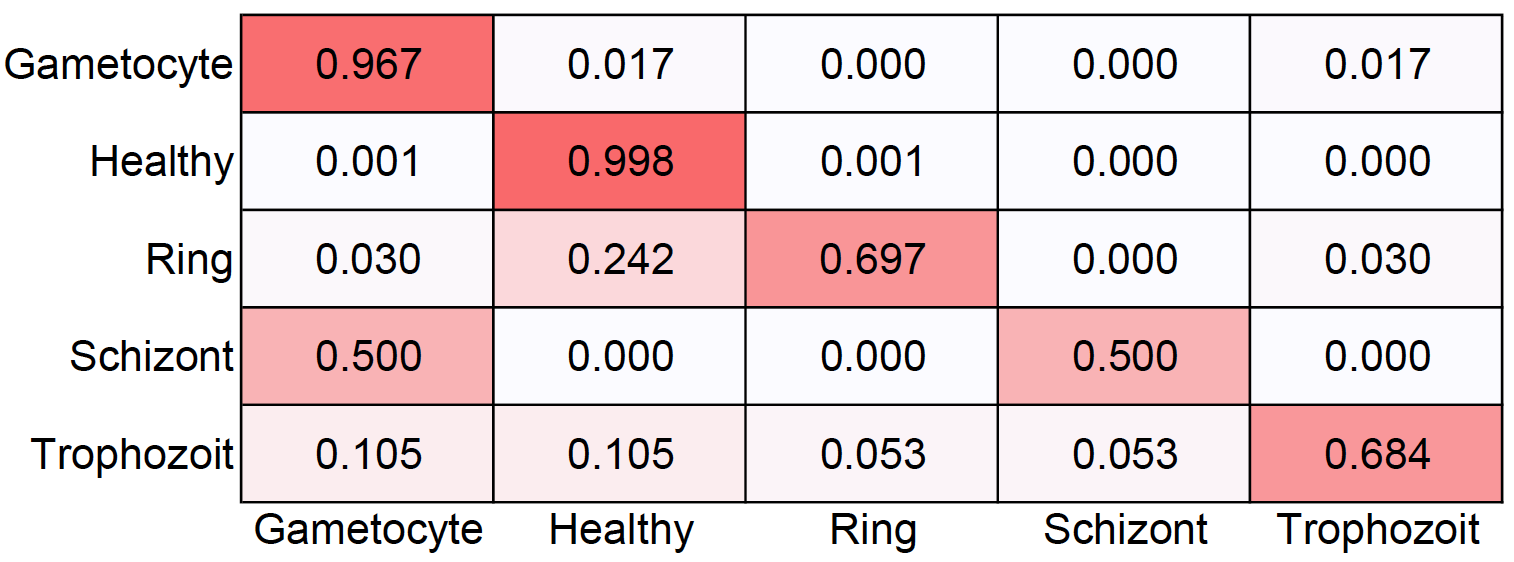}
    \caption{Two-stage multi-class classification confusion matrix which use ResNet50v2 at both stages.}
    \label{fig:two_stage_confusion_matrix}
\end{figure}

\noindent\textbf{Two-Stage Classification}
Both average accuracy and $F_1$ score improves (Table.\ref{table:2stage_classification}) when two-stage model is applied over the test data. 
As indicated by the confusion matrix, Figure \ref{fig:two_stage_confusion_matrix}, the prediction accuracy improves for the Gametocyte and Trophozoit, with a small improvement in detection of the Ring stage. However, Schizont does not show much improvement. 
This could be associated with a very small number of Schizont samples in the training and testing data. 

We perform extensive experiments to investigate how different design decisions affect the overall accuracy of the model. 
First-stage (binary stage), is trained using the different backbones and tested on the validation set. It was found that ResNet50v2 has higher accuracy than other backbone models. 
Keeping ResNet50v2 as the backbone for the first stage, we trained our second stage with different backbone networks. 
Table \ref{table:binary_classification} shows the accuracies of binary classifications and the model with ResNet50v2 backbone shows the best accuracy on the test data too. 
 
Table.\ref{table:2stage_classification} shows the result of two-stage classification on our malaria dataset. 
The combination of ResNet50v2 at stage one and stage two shows the top average accuracy of 79.61\%. Table.\ref{table:comparison_of_single_and} shows the comparison of single-stage and two-stage classification. From Table.\ref{table:comparison_of_single_and}, it can be seen that the accuracy of each network is improved in two-stage settings.
Looking at the results in the context of Table \ref{table:binary_classification}, one can see the benefit of using the first stage to balance out the samples coming to the second stage for evaluations.

\begin{table}[t]\centering

\begin{tabular}{|l|l|l|}
\hline
                    & Avg. Accuracy   & $F_1$-score         \\ \hline
Rajaraman et al. \cite{rajaraman2018pre}  & 84.93\%          & 76.27\%         \\ \hline
VGG-16             & 83.66\%          & 75.23\%         \\ \hline
VGG-19             & 94.79\%          & 90.12\%         \\ \hline
\textbf{ResNet50v2} & \textbf{95.63\%} & \textbf{89.91\%} \\ \hline
DenseNet121         & 94.77\%          & 88.60\%         \\ \hline
DenseNet169         & 95.61\%          & 89.16\%         \\ \hline
DenseNet201         & 95.18\%          & 89.45\%         \\ \hline
\end{tabular}
\caption{Classification results of stage-1 for different CNN models on  IML-Malaria dataset. ResNet50v2 shows the best accuracy for binary classification.}
  \label{table:binary_classification}
\end{table}

\begin{table}[t]\centering

\begin{tabular}{|l|c|c|}
\hline
                    & Avg. Accuracy   & $F_1$-score          \\ \hline
Rajaraman et al. \cite{rajaraman2018pre}    & 59.03\%          & 60.59\%          \\ \hline
VGG-16             & 79.04\%         & 80.72\%          \\ \hline
VGG-19             & 71.1\%            & 76.05\%         \\ \hline
\textbf{ResNet50v2} & \textbf{79.61\%} & \textbf{82.04\%} \\ \hline
DenseNet121         & 78.11\%          & 79.40\%          \\ \hline
DenseNet169         & 76.35\%          & 79.08\%          \\ \hline
DenseNet201         & 77.24\%          & 79.27\%           \\ \hline
\end{tabular}
\caption{Two-stage multi-class classification results on IML-Malaria dataset. ResNet50v2 is used in the first stage for binary classification and then all  CNN models are used for multi-class classification. Combination of ResNet50v2 at both stages shows the highest accuracy.}
  \label{table:2stage_classification}
\end{table}

\begin{table}[t]\centering
\begin{tabular}{|l|c|c|}
\hline
                & \multicolumn{1}{c|}{\begin{tabular}[c]{@{}c@{}}Avg. Accuracy\\ Single Stage\end{tabular}} & \multicolumn{1}{c|}{\begin{tabular}[c]{@{}c@{}}Avg. Accuracy\\ Two-Stage\end{tabular}} \\ \hline
Rajaraman et al. \cite{rajaraman2018pre} & 46.1\%  & \textbf{59.03\%} \\ \hline
VGG-16         & 55.35\% & \textbf{79.04\%} \\ \hline
VGG-19         & 70.44\% & \textbf{71.1\%} \\ \hline
ResNet50v2      & 72.86\% & \textbf{79.61\%} \\ \hline
DenseNet169     & 73.91\% & \textbf{76.35\%} \\ \hline
DenseNet201     & 74.56\% & \textbf{77.24\%}  \\ \hline
\end{tabular}
\caption{Comparison of single stage and two-stage classification:
  For all CNN models, two-stage classification outperforms one-stage classification accuracies.} 
\label{table:comparison_of_single_and}
\end{table}

\begin{figure}[!ht]
    \includegraphics[width=0.49\textwidth]{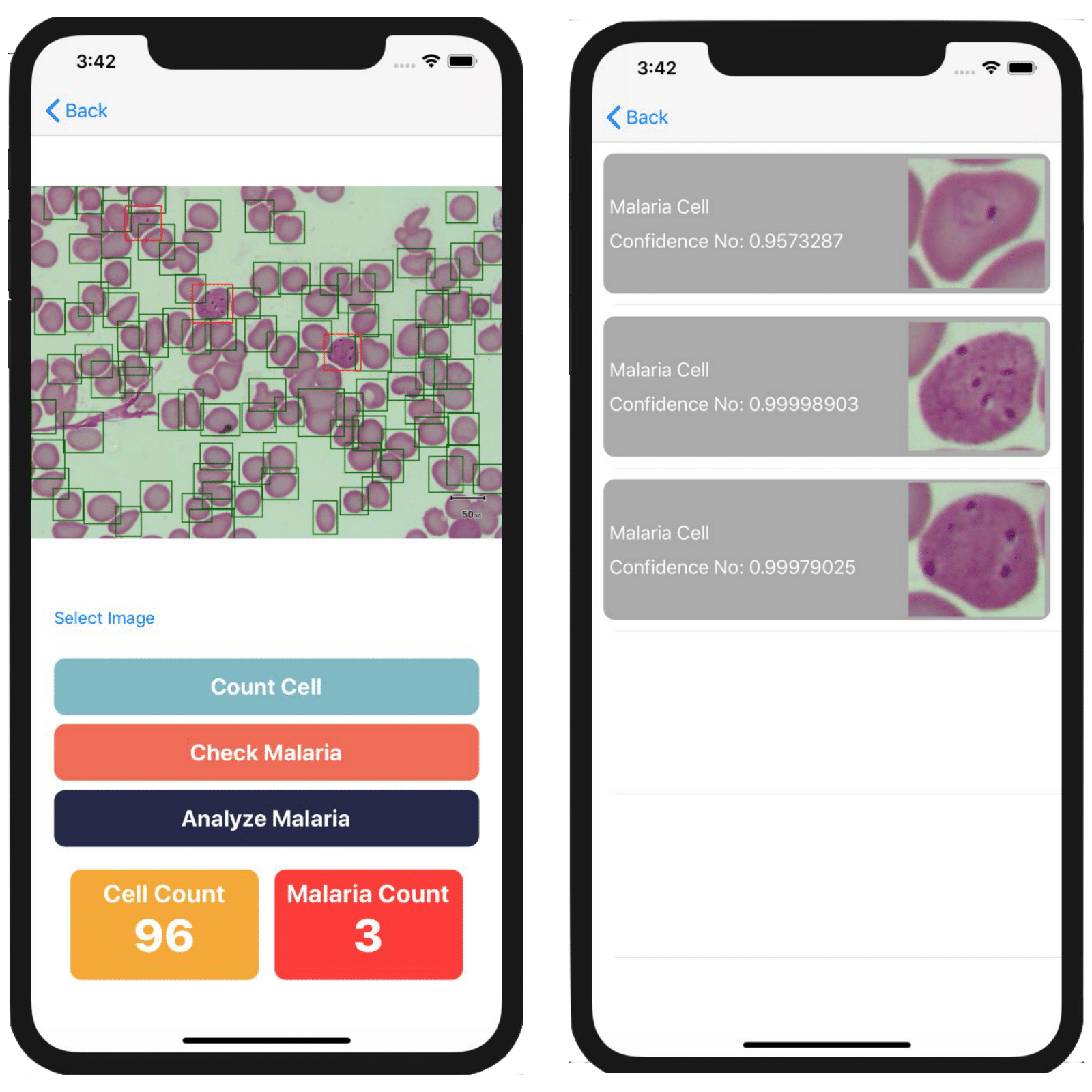}
    \caption{ Snapshot of our mobile application through one can efficiently count and localize total and  malaria infected cell in an image captured through a mobile camera mounted over the microscope.}
    \label{fig:mobile_app}
\end{figure}

\section{Mobile Application}
 
To make our solution accessible and usable in practical scenarios, we have developed a mobile application. The mobile app provides a user-friendly interface to capture images from a microscope using a mobile camera and achieve the total cell counts, number of malaria cells, and visualization of each malaria cells on the fly. Note that all processing related to CNN is done on the server and the mobile app is just used as an interface. To use the App in the field, the user just needs to connect his mobile to the internet.   

\begin{figure*}
\includegraphics[width=1\textwidth]{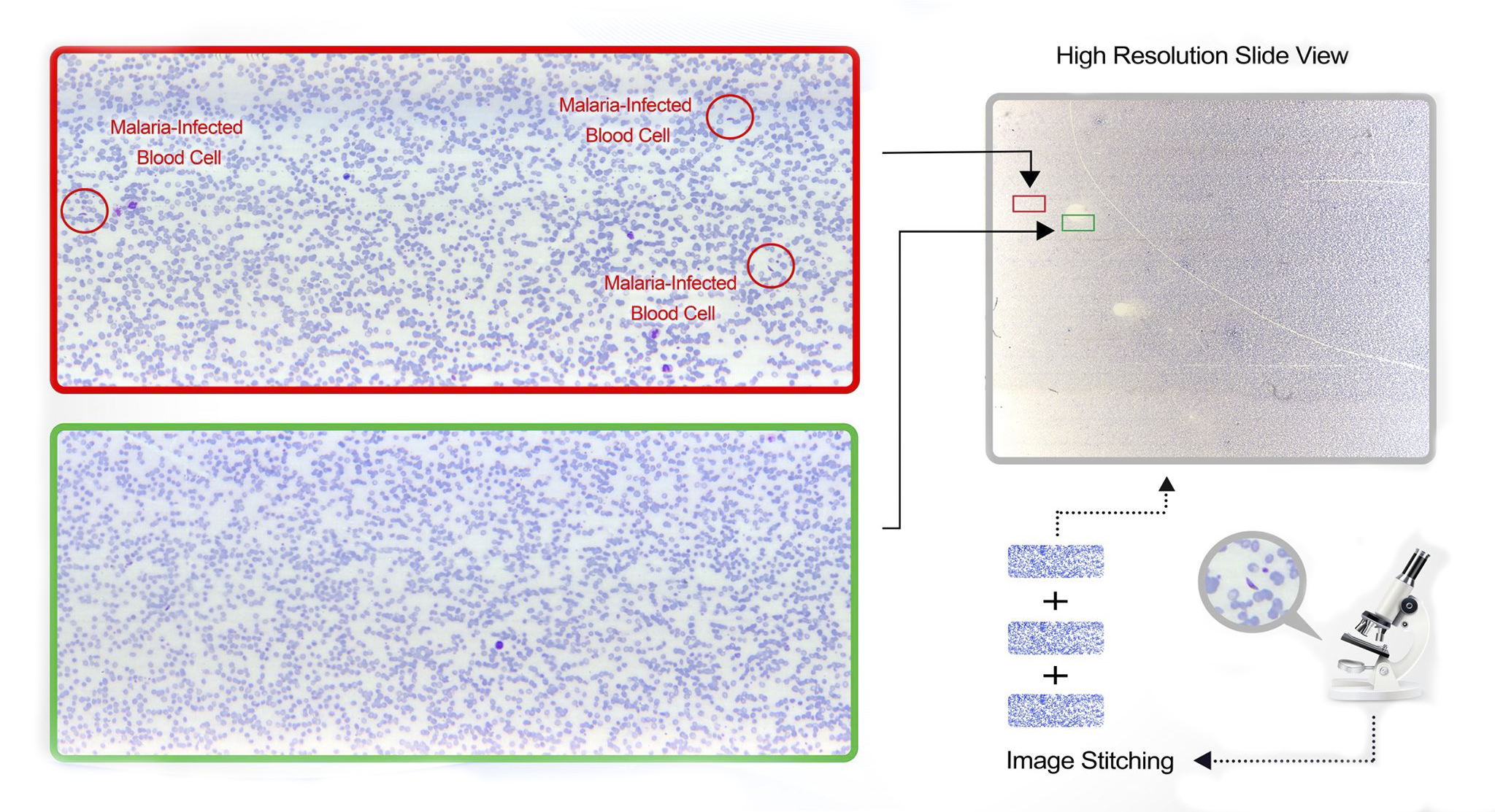}
 \caption{Blood slide image stitching that give a global view of the whole slide. Red circles shows the malaria infected blood cells. } \label{fig:stitching}
\end{figure*}

The details of the mobile app and malaria detection process through it is as follows: The mobile app is developed in Swift-5 using the Xcode IDE. Storyboard is used for app design and Alamofire API \cite{alamofire} is employed for \textit{http} networking. Server APIs are built using the Django REST framework \cite{django} that connects mobile apps with python server. 
Images are captured through a smartphone attached to a conventional light microscope. Then, the mobile app sends the image to our server that contains segmentation and classification code for cell counting, detection of malaria-infected blood cells, and classification of the malaria life cycle stage. All of these results are shown to users on mobile phones. This type of application will help many field technicians to save their time and efforts mostly in the regions where malaria is endemic.  
Our application provides a complete and end-to-end solution for detecting malaria with conventional light microscopes. Figure. \ref{fig:mobile_app} shows a brief overview of our mobile application.  

\section{Educational Use-case}
Constructing an interface for easy acquisition of the microscopic images provides us an opportunity to represent a complete slide as one. 
Manual slide examination required mechanical movement and focus adjustment repeatedly which is a very hectic task.
We implement a stitching mechanism, generating a global slide view as shown in Figure \ref{fig:stitching}.
Such representation is not only helpful for the practitioners but can be vital in educating the medical students and retraining the professionals. Training the medical practitioners and students on such recent technology will make them comfortable in using the technology to augment their expertise. This module could be extended to microscopic analysis other than malaria too.

\section{Conclusion}
In this paper, we provide a machine learning-based mechanism for automatic malaria microscopy. 
Malaria results in four hundred and five thousand fatalities every year from two hundred and twenty-eight million infected, many from the poor regions of the world. 
Machine learning-based intervention is necessary to counter the low-quality equipment and less-trained staff being used for diagnosis through blood slides and increase the speed of diagnosis since timely diagnosis is vital for a speedy recovery. 
We have collected a new large dataset of malaria microscopic images which help training deep learning-based research on this problem. 
A Convolutional Neural Network-based, two-stage pipeline is proposed for accurate malaria cell stage recognition. The cascaded nature of our architecture helps counter the effects of the imbalanced nature of the dataset.
To keep the computational cost low, instead of the deep learning-based method, a watershed algorithm is used with the morphological operations to localize the cells. 
Extensive experiments were performed to measure performance. The mean accuracy of the selected model was $79.61\%$ accuracy and $82.04\%$.
To make our approach to be easily used by medical practitioners, a new user-friendly mobile-based application is built which helps in detecting and localizing malaria cells in real-time. 
Our experimental results show that the two-stage approach performs better than the one-stage approach for the malaria life cycle stage classification. We are in the process of employing the mobile app on a large scale to verify its efficacy. \newline
\noindent\textbf{Declaration of Competing Interest:} 
We wish to confirm that there are no known conflicts of interest associated with
this publication and there has been no significant financial support for this work that
could have influenced its outcome. We confirm that the manuscript has been read and
approved by all named authors and that there are no other persons who satisfied the
criteria for authorship but are not listed. We further confirm that the order of authors
listed in the manuscript has been approved by all of us. We confirm that we have given
due consideration to the protection of intellectual property associated with this work
and that there are no impediments to publication, including the timing of publication,
with respect to intellectual property. In so doing we confirm that we have followed the
regulations of our institutions concerning intellectual property. We understand that the
Corresponding Author is the sole contact for the Editorial process (including Editorial
Manager and direct communications with the office). He/she is responsible for communicating
with the other authors about progress, submissions of revisions, and final
approval of proofs. We confirm that we have provided a current, correct email address
which is accessible by the Corresponding Author. \newline
\noindent\textbf{Acknowledgement}
 The project is partially supported by an unrestricted gift award from Facebook, USA. The
opinions, findings, and conclusions or recommendations expressed in this publication are those of the author(s) and do
not necessarily reflect those of Facebook.
 
\bibliographystyle{spmpsci} 
\bibliography{template}  

\end{document}